\documentclass[%
 reprint,
 amsmath,amssymb,
 aps,
]{revtex4-1}

\usepackage{bm}
\usepackage{graphicx}%
\usepackage{color}%

\emergencystretch=\maxdimen
\begin{document}

\preprint{APS/123-QED}

\title{Quantum phase transitions of multi-species Dirac fermions}%

\author{H.-M. Guo$^{1}$}
\author{Lei Wang$^{2}$}
\author{R. T. Scalettar$^{3}$}

\affiliation{$^1$Department of Physics, Key Laboratory of Micro-Nano Measurement-Manipulation and Physics (Ministry of Education), Beihang University, Beijing, 100191, China}
\affiliation{$^2$Beijing National Lab for Condensed Matter Physics and Institute of Physics,
Chinese Academy of Sciences, Beijing 100190, China}
\affiliation{$^3$Physics Department, University of California, Davis,
Ca 95616, USA}



\begin{abstract}
We use the determinant Quantum Monte Carlo method (DQMC) to study the
interaction-driven semimetal to antiferromagnetic insulator transition
in a $\pi$-flux Hamiltonian with modulated hoppings, a model which has
two species of Dirac fermions.  It is found that the critical
interaction strength $U_c$ is decreased by reducing the velocity of the
outer Dirac cone, while the inner cone velocity fixes the band width.
Although $U_c$ is monotonic, at fixed inverse temperature $\beta$ the
antiferromagnetic (AF) structure factor has a maximum as a function of the
hopping modulation.  We also study the corresponding strong coupling
(Heisenberg model) limit, where the sublattice magnetization is enhanced
by the alternation of the exchange couplings.  The AF order is shown to
be non-monotonic, and maximal at an intermediate degree of anisotropy,
in qualitative agreement with the Hubbard model.  These results quantify
the effect of the velocities on the critical interaction strength in
Dirac fermion systems and enable an additional degree of control which
will be useful in studying strong correlation physics in ultracold atoms
in optical lattices.
\end{abstract}

\maketitle

\textit{Introduction.-}
Much recent progress has been made in studying condensed matter emergent
quasiparticles which have close analogs in high-energy physics. A
primary example is the behavior of electrons on graphene's honeycomb
lattice, which are described by two-dimensional Dirac fermions, and
whose unusual physical properties have triggered great
interest\cite{graphene}. Dirac fermions are also central to the rapidly
developing field of topological insulators (TI). The low-energy theory
of TIs is the Dirac equation with a topological mass and an odd number
of Dirac fermions on the surface\cite{ti1,ti2,ti3,shen}.  Other
solutions of the Dirac equation, Majorana fermions (neutral particles
that are their own antiparticle) and Weyl fermions, have also been
detected in condensed-matter systems \cite{Alicea,Franz,weyl}.

These three kinds of fermions have half-integer spin, similar to their high
energy analogs, but solid state systems may also have other
quasiparticles.  Fermions described by a simple ${\bf k}\cdot
{\bf S}$ Hamiltonian with ${\bf S}$ the spin operator (obeying the Lie
algebra of SU(2)] of spin-$1$ or spin-$3/2$ have been reported, and
nearly $40$ candidate materials have been proposed
\cite{bernevig,stern,ezawa}. While these studies focus on three
spatial dimensions, realizations similar to graphene in which
fermionic quasiparticles are constrained to move in two spatial
dimensions, have been considered. The low-energy excitations of itinerate
electrons on a Lieb lattice are described by a 2D ${\bf k}\cdot {\bf S}$
Hamiltonian with pseudospin $1$, and have been realized
in photonic lattices\cite{lieb1,lieb2} and engineered atomic lattices
\cite{lieb3,lieb4,lieb5}.  Generalizations to arbitrary spin $S$ have been
obtained in stacked triangular lattices\cite{triangle}
and 2D optical superlattices\cite{lan1,lan2}.

\begin{figure*}[htbp]
\centering \includegraphics[width=10cm]{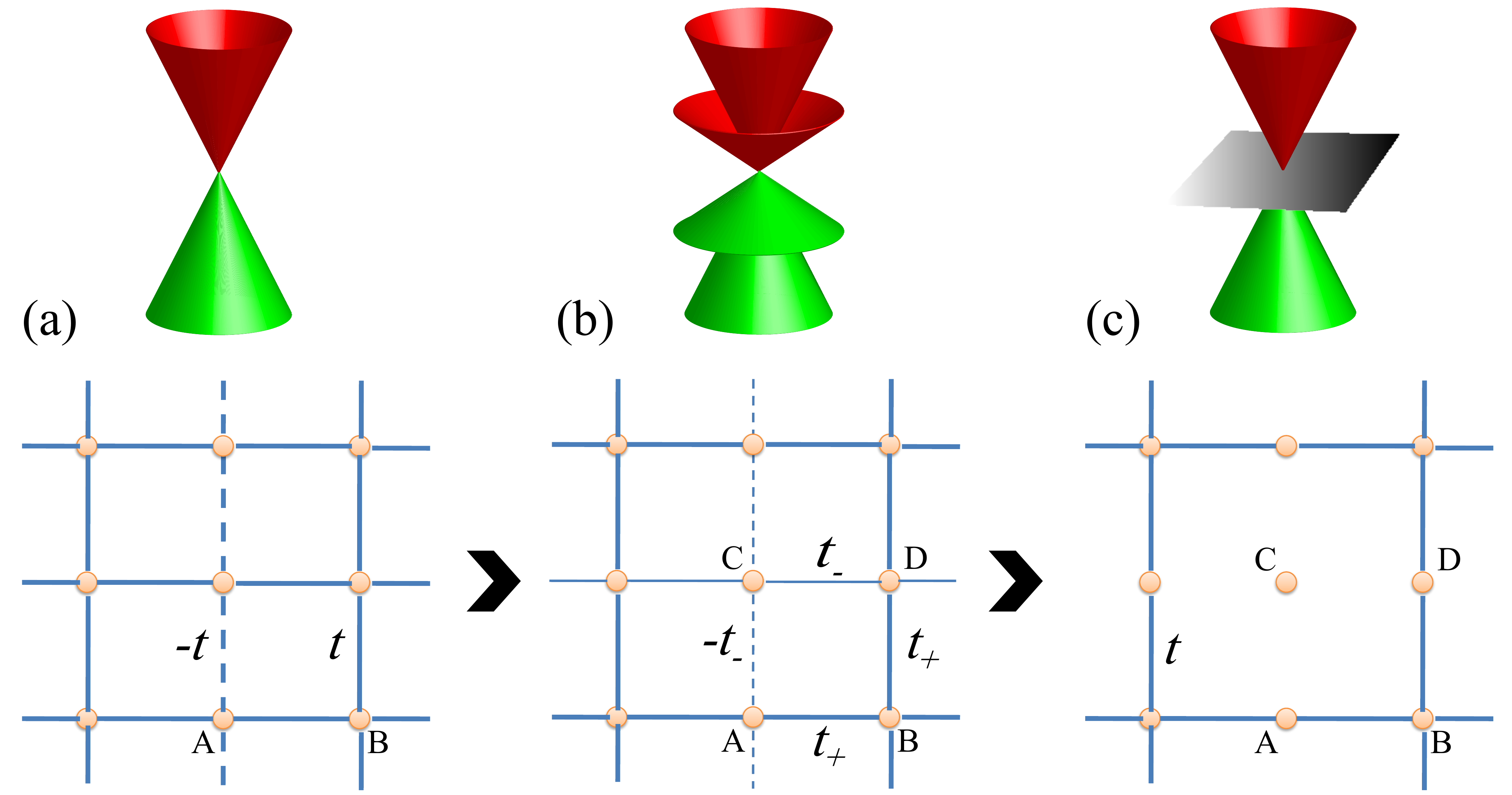}
\caption{The energy
spectra and lattice geometries considered in this paper.
(a) The $\pi$-flux lattice has two identical single Dirac cones.
The solid (dashed) lines represent positive (negative) hoppings.
(b) The unit cell is doubled by an alternation of
the hoppings $t_{\pm}=(1\pm\alpha)t_{0}$, and, as a result, there are two Dirac
cones with different velocities.
(c)
In the limit $\alpha=1$, one of the four sites in one unit
cell is completely depleted since the hoppings to it, $t_-=0$.
The resulting geometry is a Lieb lattice
($\frac{1}{4}$ depleted square lattice). }
\label{fig1}
\end{figure*}

An important phenomenon exhibited by these fermions is the
interaction-driven metal-insulator transition (MIT). Intense research
has been carried out on 2D spin-$1/2$ Dirac fermions in the Hubbard model on
a honeycomb lattice and $\pi$-flux square lattice, where the simplest
scenario of a direct and continuous transition has been
confirmed\cite{Meng,sorella1,sorella2,assaad,leiwang,zixiangli}. The
quantum critical behavior in the vicinity of the phase transition is
universal and is described by the Gross-Neveu model.  Studies of
spin-$1$ Dirac fermions have an even longer history.  The Lieb lattice
hosts such fermions and belongs to a class of bipartite geometries where
a rigorous theorem implies a ground state at half filling with a nonzero
spin and long-ranged ferromagnetic order for $U>0$\cite{lieb,scalettar}.

It is natural to study the interaction effects in other 2D spin-$S$
Dirac systems.  In this manuscript, we consider the MIT and magnetic
order in a system similar to that of spin-$3/2$ Dirac fermions in which
a birefringent breakup of the doubly degenerate yields Dirac cones with
two different `speeds of light'\cite{kennett1,kennett2,kennett3}. The
setup can be realized in the $\pi$-flux model with modulated hoppings
which we solve using DQMC. It is found that the critical interaction scales
with the velocity of the outer Dirac cone while the inner cone fixes the
band width. The present setup provides an ideal system to study the
critical interaction of Dirac fermions with continuously tuned velocity.
We also show that in the corresponding Heisenberg model, which is the
strong coupling limit of the $\pi$-flux Hubbard model, the sublattice
magnetization shows a peak as the velocity is decreased, providing an
illustration of the interesting phenomenon of enhancing quantum
antiferromagnetism by weakening the bonds.

\textit{Model and Method.-}
We start from the $\pi$-flux Hubbard model,
\begin{eqnarray}\label{eq1}
\hat{\cal H}&=&\sum_{\langle ij \rangle \sigma}t_{ij}e^{i\chi_{ij}}c^\dag_{j\sigma}c_{i\sigma}\\ \nonumber
&+&U\sum_{i}(n_{i\uparrow}-\frac{1}{2})(n_{i\downarrow}-\frac{1}{2}).
\end{eqnarray}
The noninteracting part is a tight-binding Hamiltonian on a square
lattice with each plaquette threaded with half a flux quantum,
$\frac{1}{2}\Phi_{0} =\frac{1}{2} \frac{hc}{e}$ \cite{guo,franz1}. Here
$c^\dag_{i\sigma}$ and $c^{\phantom{\dagger}}_{i\sigma}$ are
the creation and annihilation
operators at site ${\bf r}_i$ with spin $\sigma=\uparrow, \downarrow$;
the hopping amplitude between the nearest-neighbor sites $i$ and $j$ is
$t_{ij}$.  $\chi_{ij}$ is the Peierls phase, given by
$\chi_{i,i+\hat{x}}=0, \chi_{i,i+\hat{y}}=\pi \, i_x$ in the Landau gauge.
The resulting hopping pattern is shown in Fig.~1(a).  When $t_{ij}$ is
uniform, the lattice has a two-site unit cell (with labels A,B) and in reciprocal
space, with the reduced Brillouin zone $(|k_x| \leq \pi/2, |k_y| \leq
\pi)$, the noninteracting Hamiltonian can be written as
\begin{align}
H_0 &=\sum_{\bf{k}\sigma}
\psi^{\dagger}_{\bf{k}\sigma}
{\cal H}_0({\bf k})
\psi_{\bf{k}\sigma}
\hskip0.30in
\setlength{\arraycolsep}{1pt}
\psi_{\bf{k}\sigma} = \left( \begin{array}{cc}
c^{\phantom{\dagger}}_{A\sigma} &
c^{\phantom{\dagger}}_{B\sigma} \\
\end{array} \right)^{T}
\nonumber \\
& {\cal H}_0({\bf k}) =
\left( \begin{array}{cc}
 -2t \, {\rm cos} k_y & +2t \, {\rm cos}k_x \\
 +2t \, {\rm cos} k_x & +2t \, {\rm cos}k_y \\
\end{array} \right)
\end{align}
The energy spectrum $E_{\bf k} = \pm
\sqrt{4t^{2}(\cos^2 k_x+\cos^2 k_y)}$ describes a semi-metal with
two inequivalent Dirac points at ${\bf K}_{\pm}=(\pi/2,\pm \pi/2)$.

Introducing $t_{ij}$ with the pattern shown in Fig.~1(b), the unit cell
is also doubled along the $y$-direction. The two Dirac points are folded to the
same point $(\pi/2,\pi/2)$ in the reduced Brillouin zone. The
Hamiltonian in momentum space becomes,
\begin{align}
H_0 =\sum_{\bf{k}\sigma}
& \psi^{\dagger}_{\bf{k}\sigma}
{\cal H}_0({\bf k})
\psi_{\bf{k}\sigma}
\hskip0.25in
\psi_{\bf{k}\sigma} = \left(
\setlength{\arraycolsep}{1pt}
\begin{array}{cccc}
c^{\phantom{\dagger}}_{A\sigma} &
c^{\phantom{\dagger}}_{B\sigma} &
c^{\phantom{\dagger}}_{C\sigma} &
c^{\phantom{\dagger}}_{D\sigma} \\
\end{array}
\right)^{T}
\nonumber \\
{\cal H}_0({\bf k}) =
&\left(
\begin{array}{cccc}
0 & 2t_{+}\cos k_x & -2t_{-}\cos k_y & 0 \\
   & 0 & 0& 2t_{+}\cos k_y \\
  &  & 0 &  2t_{-}\cos k_x \\
  &  &  & 0 \\
\end{array} \right)
\end{align}
where the lower triangle is filled so that the
matrix is Hermitian.
Now the energy spectrum is $ E_{\bf k}=\pm
2t_{\pm}\sqrt{\cos^2 k_x+\cos^2 k_y}$ with $t_{\pm}=(1\pm\alpha)t_{0}$ and
$t_{+}=1$ defines the energy scale.  We obtain
`birefringent' fermionic Dirac
species with distinct velocities $ 2t_{\pm}$.

When treated within mean field theory,
AF order is represented by an additional real-space term
$H_{AF}=m\,\sum_{i}(-1)^{l}
(c^{\dagger}_{i\uparrow}c^{\phantom{\dagger}}_{i\uparrow}-
c^{\dagger}_{i\downarrow}c^{\phantom{\dagger}}_{i\downarrow})$,
where $(-1)^{l}=+1(-1)$ if site $l$ is on the two
sublattices of the bipartite square lattice. The spectrum
$E^\prime_{\bf k}=\pm 2t_{\pm}\sqrt{\cos^2 k_x+\cos^2 k_y+m^2}$ is
immediately gapped.

We will, instead, analyze the behavior within an exact treatment of
the correlations by simulating the
$\pi$-flux Hubbard model Eq.~(1) using the DQMC
method\cite{dqmc0,dqmc1,dqmc2,dqmc3}. To characterize the magnetic
properties, we measure the staggered structure factor,
$S(\pi,\pi)=\frac{1}{N}\sum_{i,j}(-1)^{l}\langle {\bf S}_i \cdot {\bf
S}_j\rangle$.  A physical quantity of central interest is the sublattice
magnetization, which is given by $M^2=S(\pi,\pi)/N$.  We also employ the
stochastic series expansion (SSE) QMC method with directed loop update
to simulate the corresponding Heisenberg model of the Hamiltonian Eq.(1)
in the strong coupling limit\cite{sandvik}. The SSE QMC is performed
using the ALPS libraries\cite{alps}.

\begin{figure}[htbp]
\centering \includegraphics[width=6cm]{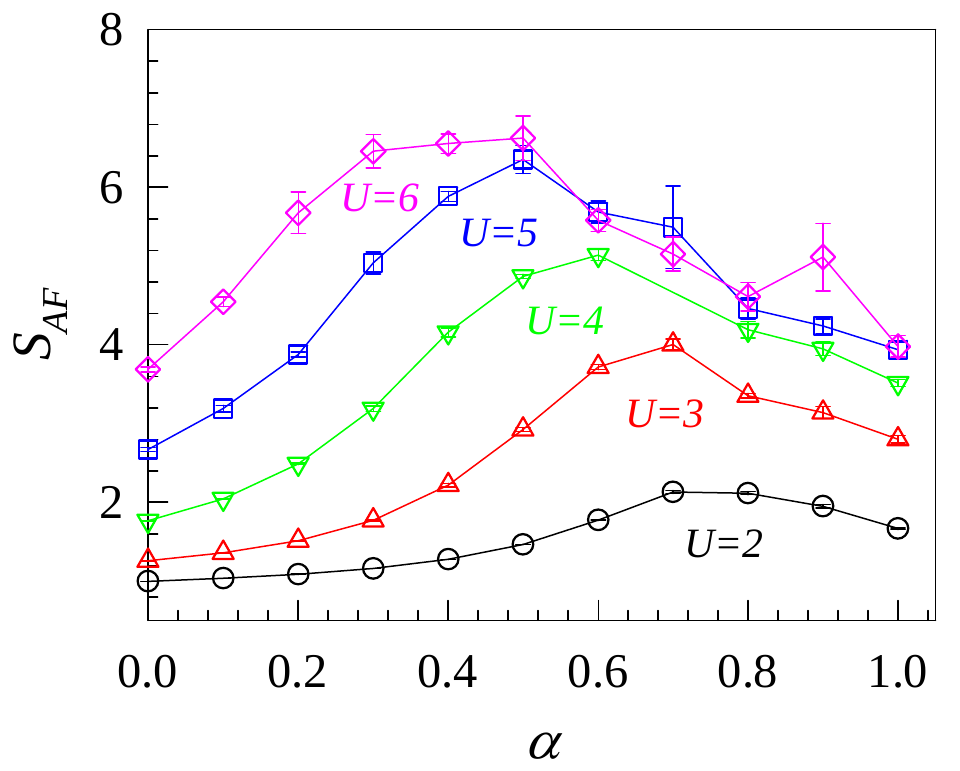}
\caption{ The AF structure factor $S_{AF}$ as a function of $\alpha$ for
different $U$ at $\beta=10$. The linear size $L=8$ and the number of sites $N=64$.
Data were acquired from $10$ simulations of $1000$ equilibration and
$10000$ measurement sweeps for each $\alpha$. }
\label{fig1}
\end{figure}

\textit{Decreased Critical Interaction at the Semimetal to AF
Insulator Transition-} Figure 2 shows $S_{AF}$ on a $N=8\times 8$
lattice for different $U$ as a function of $\alpha$. At the $\pi-$flux
lattice point ($\alpha=0$), it is known that long-ranged
antiferromagnetic order (LRAFO) exists when $U$ exceeds
$U_c=5.65\pm
0.05$\cite{sorella2}. In the $\alpha=1$ limit, the geometry is the Lieb
Lattice, where LRAFO (or, more precisely, ferrimagnetic order)
exists for all $U>0$.
The behavior of $S_{AF}$ is qualitatively similar for different values of the
interaction strength $U$: $S_{AF}$ first increases
at small $\alpha$; quantum
antiferromagnetism is enhanced.
However, after reaching a maximum at intermediate $0<\alpha <1$,
the structure factor falls off.
Intuitively, the final decrease in spin-spin
correlations as $\alpha \rightarrow 1$ might be associated
with the bonds are being increasingly weakened, and indeed are finally
completely depleted from the lattice at $\alpha=1$.

\begin{figure}[htbp]
\centering \includegraphics[width=7.5cm]{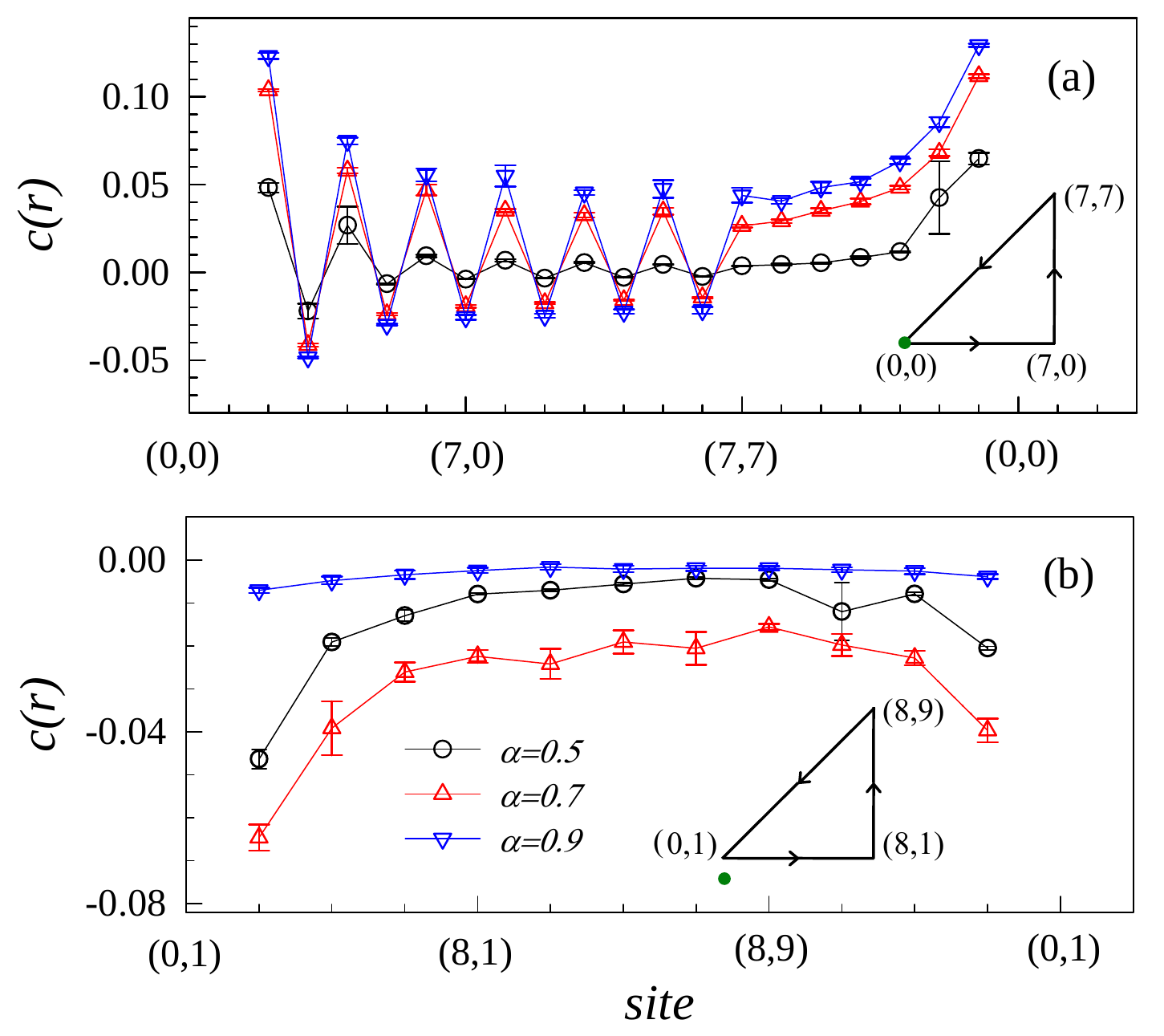} \caption{The equal
time spin-spin correlation function $c({\bf r})$ on a $16\times 16$
lattice at $\beta=10$. Data are shown for various $\alpha$ at $U=3$, including
$\alpha=0.7$ which has the largest $S_{AF}$. (a), $c({\bf r})$ between
$A$ site at $(0,0)$ and $A, B, D$ sites along the triangular path on the
lattice shown in the inset. (b), $c({\bf r})$ between $A$ site at
$(0,0)$ and $C$ sites on the path of the inset. The green dot denotes the $A$ site at the origin $(0,0)$. }
\label{fig1}
\end{figure}

To gain additional insight into the behavior of $S_{AF}$, it is useful
to examine the
equal-time {\it real space} spin-spin correlation function
$c({\bf r})=\langle (n_{j+{\bf r}\uparrow}-n_{j+{\bf
r}\downarrow})(n_{j\uparrow}-n_{j\downarrow})\rangle$.
$c({\bf r})$ is spin-rotation invariant, and in our simulations
we average the $z$ correlation above over all three
directions to provide an improved estimator\cite{scalettar1}.
Figure 3 shows the spin-spin correlation
function for $U=3$ at $\alpha=0.5, 0.7, 0.9$ on a $16\times 16$ lattice.
The origin is placed on the $A$ site at $(0,0)$, and ${\bf r}$ runs
along a triangle path. The path in Fig.~3(a) contains $A, B, D$ sites.
The absolute values of $c({\bf r})$ grow monotonically as $\alpha$ is
increased, reaching the Lieb lattice limit at
$\alpha=1$. Their signs are consistent with antiferromagnetic order.
$\alpha=0.5$ is close to the critical point, thus the value
of the correlation at large distance is relatively small at the
temperature considered.  On the other hand,
the correlation length has become
comparable to the system size for the cases of $\alpha=0.7, 0.9$.
$c({\bf r})$ has a robust persistence at large distance, suggesting the
existence of the LRAFO.

Along the path in Fig.~3(b), which only contains
$C$ sites, the correlations are always negative since the sites
are on the opposite sublattice from the origin.
The trend in their
amplitude with increasing $\alpha$ is initially the same as
that of Fig.~3(a):  the correlations grow in size with  $\alpha$.
Crucially, however, after reaching a
maximum at $\alpha \sim 0.7$ they rapidly decline.
This is then the real-space origin of the non-monotonicity
of $S_{AF}$ with $\alpha$.

The behavior of $S_{AF}$ is suggestive of the fact that LRAFO may
develop at a decreased critical interaction in the presence of bond
weakening in going from the $\pi$-flux lattice to the Lieb lattice.
We use finite size scaling on lattices of sizes
$L=8,12,16$ to analyze quantitatively the position
of the
critical point in the thermodynamic limit.
The square of the order parameter is given by
$S_{AF}/N$  in the limit $1/L\rightarrow 0$.
These extrapolated values are shown in Fig.~4. As a function
of $t_{-}=v/2$ ($v$ the velocity of the outer Dirac cone), the
critical interaction strength is continuously decreased to zero.

Recent studies of the honeycomb and $\pi$-flux Hubbard models have
suggested that the velocities of the Dirac fermions are the main
contribution to the renormalization of $U_c$\cite{sorella2,assaad}. However, the velocities in
these models are not tunable if one wants to keep the band widths fixed.
In contrast, in
the system studied here, the velocities of one species of Dirac fermions
are continuously tuned, while the other species fixes the band width.
Thus one is able to make a definitive statement concerning
the evolution of $U_c$ measured in the meaningful limit
of fixed bandwidth.

\begin{figure}[htbp]
\centering \includegraphics[width=7.5cm]{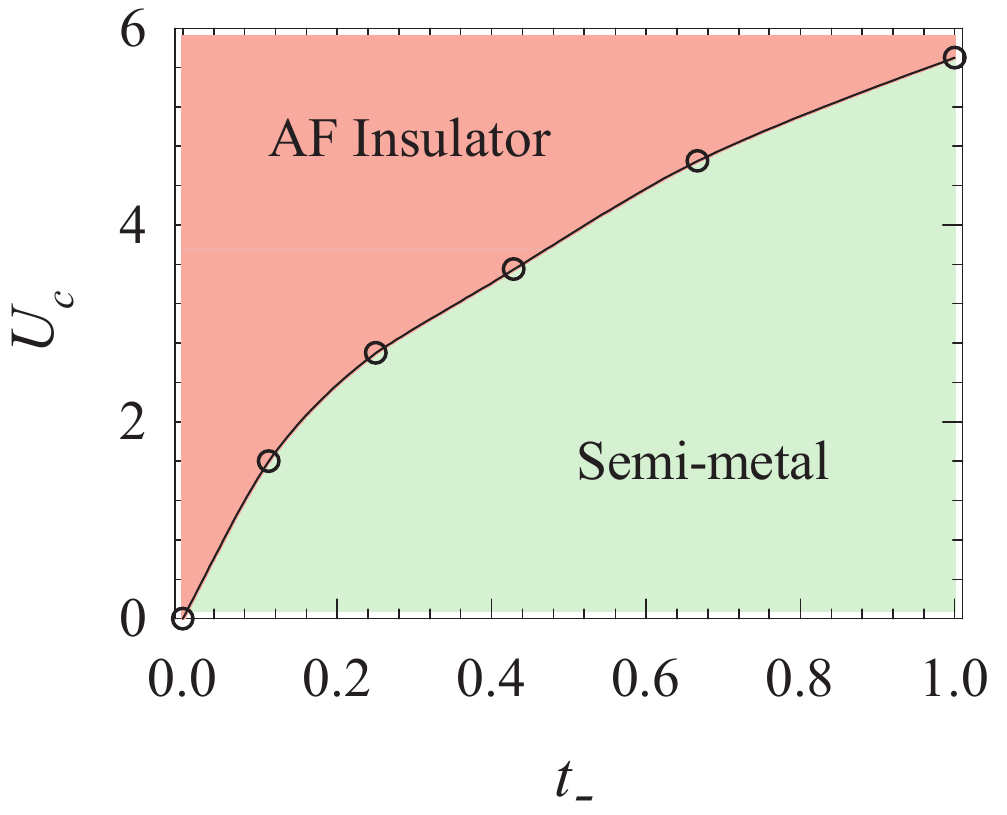} \caption{The
critical interaction $U_c$ in the thermodynamic limit as a function $t_{-}$ (the hopping amplitudes on the weakened bonds).
 The velocity of the outer Dirac cone is $v=2t_{-}$. The inverse temperature is $\beta=20$.}
\label{fig1}
\end{figure}

\textit{2D Heisenberg model with Weakened Bonds-}
It is well known that the half-filled Hubbard model, which describes itinerant
magnetism, maps
onto the antiferromagnetic (localized spin-1/2) Heisenberg model
\begin{eqnarray}\label{eq1}
H=J\sum_{\langle i,j\rangle}{\bf S}_{i}\cdot {\bf S}_{j}
\hskip0.30in (J>0),
\end{eqnarray}
as $U/t \rightarrow \infty$.  The relation
$J=4t^2/U$ gives the exchange constant in terms of $U$ and the hopping
amplitude.
The 2D square lattice Heisenberg model with uniform $J$
has been studied intensely by means of various
theoretical and numerical techniques. There is a general consensus that
antiferromagnetic long-range order exists in the ground
state\cite{reger88}.
We consider here the inhomogeneous variant
corresponding to the Hamiltonian Eq.~(1) describing two species of Dirac
fermions, in which the antiferromagnetic coupling is modulated with
$J_{\pm}=(1\pm \alpha)^2 J$.  (See inset to Fig.~5.)

\begin{figure}[t]
\centering
\centering \includegraphics[width=7.cm]{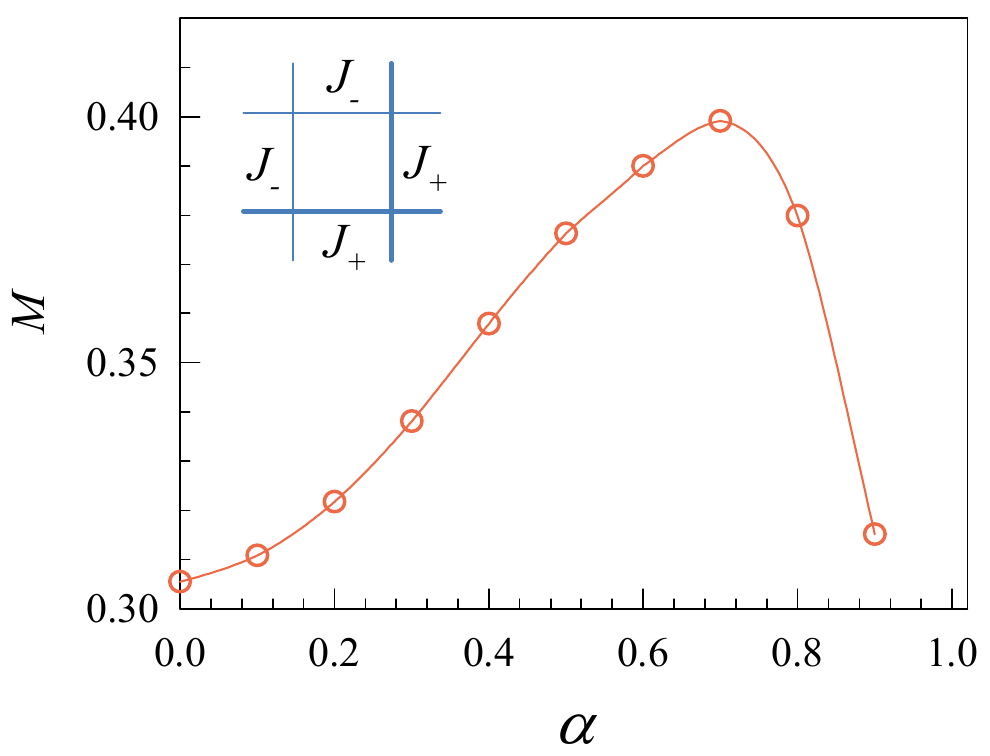} \caption{ The sublattice
magnetization of the 2D spin-1/2 Heisenberg model
in the thermodynamic limit, as a function of the modulation strength $\alpha$
of the exchange constants. The
inverse temperature is $\beta=100$, which is low enough to acquire the
ground state\cite{note}. The inset shows the four-site unit cell with the modulated
coupling $J_{\pm}$. In the simulations, $J_{+}=1$ is fixed as the energy
scale, and $J_{-}=(1-\alpha)^2/(1+\alpha)^2$ on the
weakened bonds.}
\label{fig1}
\end{figure}

$M$ is calculated on $L\times L$ lattices with $L$ up to $48$, and is
extrapolated to the thermodynamic limit using fits to polynomials in
$1/L$. Figure 5 shows the extrapolated values as a function of $\alpha$.
At $\alpha=0$ we recover the 2D Heisenberg model, obtaining
$M=0.3055\pm 0.0015$, which is consistent with previous QMC
results\cite{sandvika,sorella3}. As the bonds are weakened, the
antiferromagnetic correlations are enhanced. The data shows a peak with
a maximum $M=0.399$ at about $\alpha=0.7$.  The order parameter
increases to about $80\%$ of the classical limit, emphasizing
the reduction of quantum
fluctuations.  At this maximum, the coupling $J_{-}=0.0311$.
Beyond $\alpha \sim 0.7$, $M$
decreases quickly. In the limit $\alpha=1$, where $J_{-}=0$, the square lattice
is depleted to Lieb lattice, and the extrapolated value, $M=0.3842$.
The enhancement of the sublattice magnetization can also be explained
by analyzing
explicitly the details of the real-space
spin-spin correlation function, similar to the fermionic case
shown in Fig.~3.

There are other ways to arrange weakened couplings, generating
anisotropic, dimerized and plaquette Heisenberg models, which have been
much studied in the
literature\cite{anisotropic,plaquette,dimerized,dimerized1}. However in
these cases, the sublattice magnetization decreases monotonically as the
bonds are weakened. The quantum antiferromagnetic enhancement found
here is intriguing since it does not correspond
to the qualitative trends in this past work.

\textit{Conclusions.-}
The interaction-driven semimetal to antiferromagnetic insulator
transition of two coupled species of Dirac fermions, realized by
depleting the $\pi-$flux lattice to the Lieb lattice, was studied using
the DQMC method.  During the process of increasing the modulation of the
hopping, the quantum antiferromagnetism is enhanced, resulting in a
decreased critical interaction $U_c$ at the quantum phase transition.  A
related phenomenon of increased sublattice magnetization in the
corresponding Heisenberg model was also confirmed. The Hamiltonian
studied provides a clean, idealized system in which to explore the
effect of velocity on the critical interaction strength for magnetic
ordering and MITs of Dirac fermions.

Moreover,
ultracold atoms in optical lattices provide a platform to implement this
system. Very recently, the 2D Fermi-Hubbard model has been realized in a
series of experiments and spin correlations displaying antiferromagnetic
behavior have been observed directly with Bragg scattering\cite{hart15}
and fermionic
microscopes\cite{fhubbard1,fhubbard2,fhubbard3,fhubbard4}.  In addition, a
scheme based on resonant modulations was developed to engineer synthetic
gauge fields and a constant flux per plaquette throughout the optical
lattice\cite{gauge}.  This offers the possibility of `building'
the modified $\pi$-flux
Hubbard model whose physics was explored here,
and verifying experimentally our key findings.

The enhancement of magnetic order with inhomogeneous hopping patterns
which we find is, in fact, an intriguing feature of a
number of strong correlation phenomena.  It has been proposed,
for example, that an `optimal inhomogeneity' exists for $d$-wave
superconductivity in a Hubbard model which builds a
full 2D lattice from 2x2 plaquettes serving as binding centers
for $d$-wave pairs\cite{yao07,baruch10,fye92,smith13}.
However, a considerable degree of discussion has
arisen from different results obtained within complementary
analytic and numeric apporaches, both for the magnetic
and pairing responses\cite{doluweera08,maier06,chakraborty11}.
We have shown that the idea of optimal inhomogeneity extends
further to Dirac fermions, and also proposed a novel way to
tune the inhomogeneity so that the velocity can be used to
adjust the position of the quantum critical point
between paramagnetic metal and antiferromagnetic insulating
ground state phases.

\textit{Acknowledgments-}
The authors thank R. Mondaini, R. R. P. Singh and  Shengyuan A. Yang for helpful discussions. H.G.
acknowledges support from China Scholarship Council and NSFC grant No. 11774019. R.T.S.~was supported by DOE grant DE-SC0014671.

\nocite{*}

\bibliography{apssamp}

\begin{thebibliography}{60}
\bibitem [{not()}]{note}%
  \BibitemOpen
  \href@noop {} {\bibinfo  {journal} {Here we use $\beta=100$ for the
  Heisenberg model but only $\beta=20$ for the Hubbard one. The reason is that
  one needs to go to lower temperature on larger lattices. For the Heisenberg
  case, we use the lattice size $L=48$, thus the temperature needs to be colder
  to get the correlation length to reach the lattice size}\ }\BibitemShut
  {NoStop}%
\end{thebibliography}

\end{document}